\documentclass[letterpaper,12pt]{article}   
\usepackage[draft]{hyperref} 
\usepackage{graphicx}
\usepackage{amsmath}
\usepackage{color}
\usepackage[affil-it]{authblk}

\begin{document}

\title{Beyond the diffraction limit via optical amplification}

\author{Agla\'e N. Kellerer}
\affil{Cavendish Laboratory, Cambridge University, UK \\ \texttt{ak935@cam.ac.uk}}

\author{Erez N. Ribak }
\affil{Physics Dept., Technion - Israel Institute of Technology, Haifa 32000, Israel}


\maketitle

{\bf Abstract:\/}
In a previous article we suggested a method to overcome the diffraction limit behind a telescope. We refer to theory and recent numerical simulations, and test whether it is indeed possible to use photon amplification to enhance the angular resolution of a telescope or a microscope beyond the diffraction limit. An essential addition is the proposal to select events with above-average ratio of stimulated to spontaneous photons. We find that the diffraction limit of a telescope is surpassed by a factor ten for an amplifier gain of 200, if the analysis is restricted to a tenth of the incoming astronomical photons. A gain of 70 is sufficient with a hundredth of the photons.

\vskip 2cm

The angular resolution of telescopes is believed to be ultimately bounded by the diffraction limit. The diffraction limit has however been overcome in other domains, perhaps most notably in microscopy\,\cite{Hell}. Unfortunately, the methods developed for microscopy can not be applied to astronomy, because the observed sample needs to be illuminated with coherent laser light -- an obvious impossibility on a distant astronomical target. We have however recently suggested an approach to overcome the diffraction limit in astronomy.  The method is based on optical amplification\,\cite{Kellerer}:  the incoming astronomical photon is copied via stimulated emission and the momentum of the photon is determined from the set of stimulated photons, rather than from a single photon. Photon amplification via stimulated emission is always accompanied by spontaneous emissions. The minimum amount of spontaneous emissions has been discussed i.a.  by Milonni \& Hardies \cite{Milonni}, Wootters \& Zurek \cite{Wootters}, Mandel \cite{Mandel} and Caves \cite{Caves}: An optical amplifier emits a minimum number of spontaneous photons; this minimum is such that the Heisenberg uncertainty relation is preserved. 
Since the diffraction limit is rooted in the Heisenberg uncertainty relation -- the localization of a photon on the entrance pupil of an optical device introduces uncertainty on the photon's momentum, i.e. on its direction -- one may be tempted to conclude that optical amplification can't allow to overcome the diffraction limit.

Kurek et al. \cite{Kurek} nevertheless report improved angular resolution beyond the diffraction limit from simulations of single photon amplification and an analysis that employs the 'matched filtering method'. Here we revisit their analysis and find that the use of the actual photon statistics behind an optical amplifier eliminates the apparent gain in angular resolution. 

 \begin{figure*}
\begin{center}
\includegraphics[width=.8\textwidth]{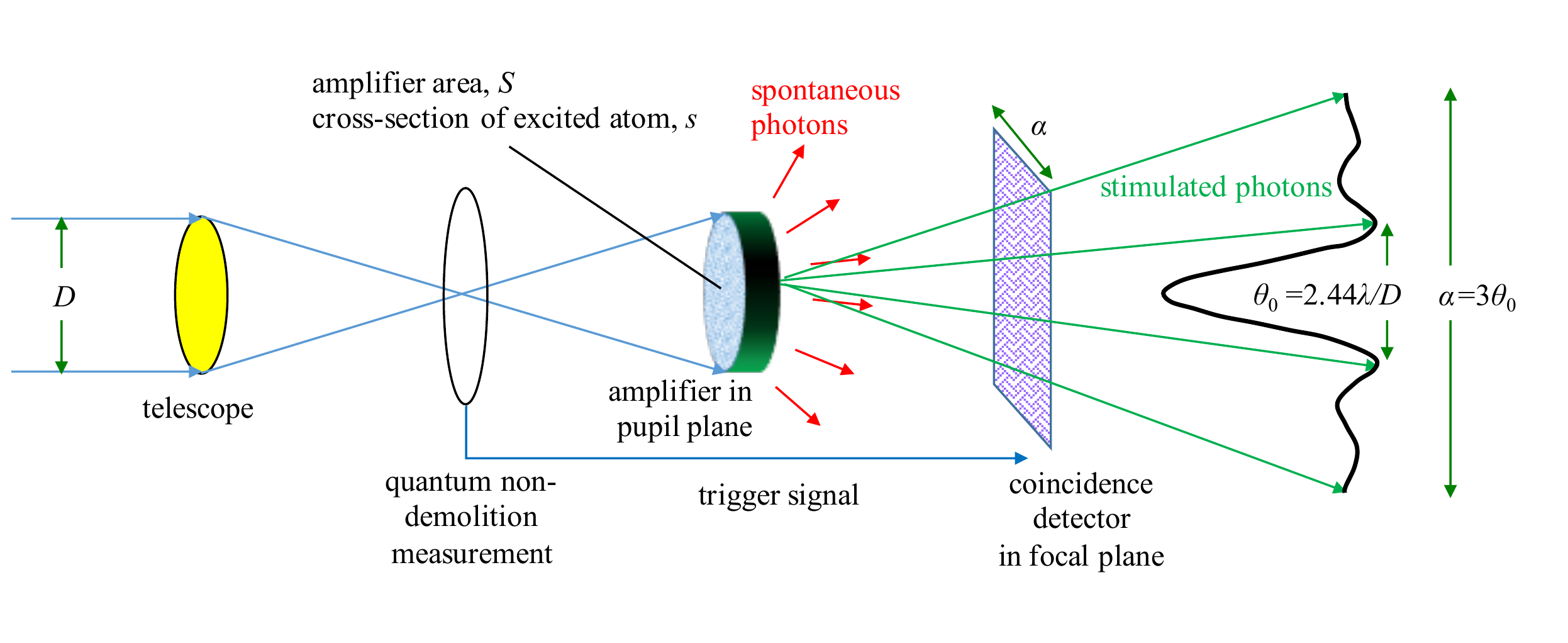}
\caption{ An amplifier medium placed behind a telescope. The surfaces of the amplifier and telescope pupil are assumed equal to simplify the notations, but in actuality the amplifier surface can be made substantially smaller than the pupil. }
\label{fig:setup}
\end{center}
\end{figure*}

An improvement in angular resolution may however be obtained by a suitable selection of events with above average ratio of stimulated to spontaneous photons. This harks back to the idea of probabilistic noiseless amplification as discussed i.e. by Duan \& Guo \cite{Duan}, Ralph \& Lund \cite{Ralph}, Zavatta et al. \cite{Zavatta} and Usuga et al. \cite{Usuga}. Probabilistic noiseless amplification allows to overcome the Heisenberg uncertainty relation on a reduced fraction of events. 

Our numerical simulations suggest an improved resolution in terms of photon amplification if the analysis is restricted to larger than average photon bursts. 
 This is an essential addition compared to Kellerer \cite{Kellerer} and Kurek et al.\cite{Kurek}, who suggest to keep all amplification events. We find that the restriction to above average size of bursts is necessary in order to overcome the diffraction limit.
 
These results may seem in contradiction with earlier results by Prasad \cite{Prasad} who states that optical amplification unacceptably corrupts astronomical signals by adding noise: ``[Amplification] adds only an insignificant amount of noise photons, provided that the average number of input signal photons per mode is large compared with 1. This is not true of optical radiation from a typical astronomical source, which is essentially a blackbody with an effective surface temperature $T$ of the order of or lower than that of the Sun, namely, 6000\,K. The average photon number per mode, $1/[\exp(hc/(\lambda\,kT)) - 1]$, at the characteristic optical wavelength of $\lambda=500$\,nm and at surface temperatures smaller than the Sun's, is much smaller than 1.''  Whilst we agree that optical amplification cannot boost astronomical signals, we argue that it may improve the angular resolution of telescopes. We consider the regime that Prasad suggests to avoid -- that is the regime where the average number of photons per mode lies below 1. As photons pass the telescope one by one, their arrival times are individually detected (non-destructively) and then each photon is amplified. The burst of amplified photons arrives on a detector, whose read-out is triggered by the quantum non-demolition detection of the astronomical photon. The incoming direction of the astronomical photon is estimated with improved precision on the burst of amplified photons. 
This article first discusses the photon statistics behind an optical amplifier and then presents numerical simulations. While we take our cue from astronomy, the results are more general and may well pertain to other cases such as microscopy.

In an amplifier medium, the mean number of stimulated photons per incoming photon is
\begin{eqnarray}
<n>= \frac{s}{S}\,I
 \end{eqnarray} 
 where $I$ is the number of excited atoms, $s$ is the cross-section of excited atoms and $S$ is the aperture- and amplifier-area, see also Fig.\,\ref{fig:setup}. A field of angular diameter   $\theta_0=2.44\,\lambda/D$ -- where $D$ is the aperture diameter and $\lambda$ the wavelength -- corresponds to the Airy disc up to its first minimum. Now that we have the number of stimulated photons, we wish to calculate the number of spontaneous photons, which are created simultaneously within the diffraction angle $\theta_0$. Within the read-out time $\Delta t=\lambda^2/(c\,\Delta\lambda)$, equal to the photon coherence time, this `diffraction volume' receives a mean number of spontaneous photons: 
 \begin{eqnarray}
<m>=\frac{\pi\,\theta_0^2}{4}\cdot\frac{I}{4\pi}\cdot A\,\Delta t
 \end{eqnarray}
 $A$ is the emission rate of spontaneous photons within $4\,\pi$ steradians: $A=8\pi\,c\,s\,\Delta\lambda/\lambda^4$\,[s$^{-1}$]. 
 
From these relations one obtains the average fluence ratio of spontaneous and stimulated photons on the diffraction area: 
\begin{eqnarray}
\frac{<m>}{<n>}=0.74\,\pi^2\sim7.3
\end{eqnarray}

The spontaneous photons within the coherence volume -- i.e. within the angular diffraction area and during the coherence time -- exceed the number of stimulated photons per incoming photon by the factor 7.3 \cite{Kellerer}. 

This level of unavoidable noise appears to rule out any improvement of resolution through amplification. However, $<n>$ and $<m>$ represent the average numbers of stimulated and spontaneous emissions. In the following, we indicate the probability distributions of the photon numbers: the probability distribution of stimulated emissions has a large variance, while the spontaneous emissions are Poisson distributed, i.e. their numbers are fairly constant. This enables to select favorable events, where the ratio of stimulated to spontaneous photons is especially large.  

The probability density of the number of stimulated photons emitted by a fully inverted amplifier medium of gain $g$ in response to a single incoming photon is given by i.a. by Shimoda et al. \cite{Shimoda} and Haus \cite{Haus}:
\begin{eqnarray}
p_{\rm st\/}(n)=\left(1-\frac{1}{g}\right)^{n}\,\frac{1}{g} \label{eq:BE}
\end{eqnarray}
which is, for reasonably large photon numbers $n$, an exponential function with mean value $<n> \sim g-1$ and with standard deviation of the photon number also equal to $\sigma_{\rm st\/}\sim g-1$.

The probability density of the number of spontaneous photons emitted in response to a single spontaneous photon is given by the same distribution (Eq.\,\ref{eq:BE}). For imaging purposes one requires a multi-mode, degenerate amplifier. As discussed by Haus \cite{Haus}, the observed spontaneous photons then originate from multiple avalanches and, in the limit of a sufficiently degenerate amplifier (many amplified modes), the probability density follows a Poisson distribution:
\begin{eqnarray}
p_{\rm sp\/}(n)=\exp(-<m>)\,\frac{<m>^n}{n!}
\label{eq:Poisson}
\end{eqnarray}
The spontaneous photons are emitted uniformly into all modes of the amplifier. The distribution has an average $<m>$ and a standard deviation $\sigma_{\rm sp\/}=\sqrt{<m>}$. For an amplifier gain, $g$, and a field diameter, $\alpha$: 
\begin{eqnarray}
<m>=7.3\,(g-1)\left(\frac{\alpha}{\theta_0}\right)^2
\end{eqnarray}
over one coherence time, $\Delta t$. We assume that the spectral width of the amplifier is matched to the spectral width of the incoming astronomical photons, so that the coherence time of the spontaneous photons equals the coherence time of the incoming, astronomical photons.

To stress the essential point: the stimulated mode exhibits super-Poissonian statistics, such that for high gain the standard deviation approximately equals the mean, whilst the spontaneous photons are spread over many incoherent modes and thus, exhibit approximately Poissonian statistics. Since the spontaneous photons are Poisson distributed, the relative standard deviation of $p_{\rm sp\/}$ decreases with increasing gain factor, $g$. This is not the case for the stimulated photons, where the relative standard deviation of $p_{\rm st\/}$ is constant with increasing gain factor. This speaks in favor of focusing on larger bursts of stimulated photons.

The stimulated photons arrive within the Airy pattern. The probability density of the angular deviation, $\theta$, from the direction of the incoming photon is
\begin{eqnarray}
f(\theta)  = \left( \frac{2 J_1(\pi\,D\,\theta/\lambda)}{\pi\,D\,\theta/\lambda}\right)^2 
=\left( \frac{2 J_1(2.44\,\pi\,\theta/\theta_0)}{2.44\,\pi\,\theta/\theta_0}\right)^2 \label{eq:Airy}
\end{eqnarray}
where $D$ is the diameter of the telescope pupil and $\lambda$ the photon wavelength.

The spontaneous photons distribute uniformly within the field-of-view of angular diameter $\alpha$.  
The average ratio of spontaneous to stimulated photons equals 7.3 within the Airy disc of angular diameter $\theta_0=2.44\lambda/D$.

Our numerical simulations assume a field of angular diameter $\alpha=3\,\theta_0$, where $\theta_0$ is the diameter of the Airy disc up to its first minimum. 
The direction of the incoming photon is randomly varied between iterations within the circular field of diameter $\alpha$.  
For a single photon the rms-error of the direction estimate equals approximately $\sigma_0\sim0.8\lambda/D$ (Eq.\,\ref{eq:Airy}).  In the classical approach, i.e without amplification and on an unresolved source, the rms-error decreases to $\sigma_0/\sqrt{n}$, if instead of a single incoming photon $n$ photons are registered.

When, on the other hand,  there is uniformly distributed noise from spontaneous emission with $7.3\,n$ photons per diffraction area and, superimposed on this noise, a burst -- to use this short term -- of $n$ stimulated photons distributed in an Airy pattern, the question arises whether within the resulting point pattern  the location of the burst of stimulated photons can nevertheless be recognized, and whether its centre might then be localized with rms-error less than $\sigma_0$.  This has recently been studied by Kurek et al. \cite{Kurek}, who find that the angular resolution can indeed be improved well beyond the diffraction limit. Kurek et al. \cite{Kurek} assume that the probability densities of the number of stimulated and spontaneous photons follow Poisson distributions.  

To test and extend Kurek et al.'s results, we performed numerical simulations in terms of different modalities:

\begin{enumerate}
\item The number $n$ of stimulated photons is taken to follow a Poisson distribution of mean value $g-1$. The number of spontaneous photons likewise follows a Poisson distribution. In a circle of angular diameter $\theta_0$ the mean number of photons equals $K= 7.3\,(g-1)$.
The standard deviation, $\sigma$, of the estimated location of the center is then determined from 2000 simulated point patterns. The Airy pattern (Eq.\,\ref{eq:Airy}) is used for the position of the stimulated photons, while the spontaneous photons are randomly distributed across the field of angular diameter $\alpha=3\,\theta_0$. The gain factor, $g$, is varied between 1 and 1000.
\item The same computations as in Mode 1, but with the actual distribution of the number, $n$, of stimulated photons: i.e. the distribution described by Eq.\,\ref{eq:BE}, rather than a Poisson distribution. Note that the simulations use the distribution described by Eq.\,\ref{eq:BE} and not its exponential approximation. 
\end{enumerate}

\begin{figure}
\centering
\includegraphics[width=.45\textwidth]{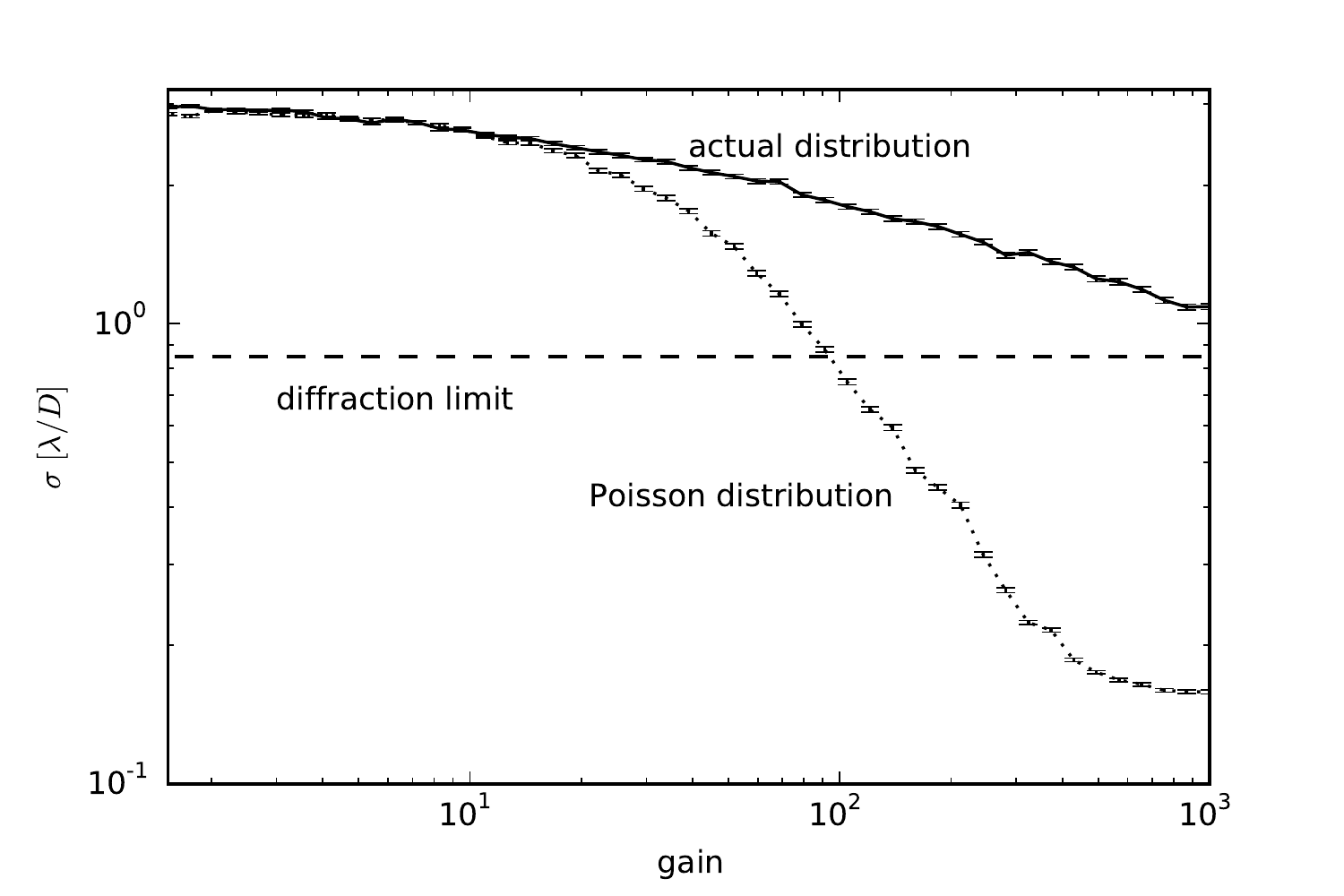}
\caption{ The rms-deviation of the position estimate. The number of stimulated photons either follow a Poisson distribution or the actual distribution indicated by Eq.\,\ref{eq:BE}: Modes 1 and 2 in the text. }
\label{fig:distr}
\end{figure}

Fig.\,\ref{fig:distr} shows that the angular resolution is significantly improved, if the numbers of stimulated photons are assumed to follow a Poisson distribution (Mode 1). This result is in agreement with Kurek et al.'s \cite{Kurek} findings. 
The actual distribution of stimulated photons behind an optical amplifier has, however, a substantially larger variance. The broad distribution of the size of the stimulated bursts enhances the standard deviation of the location estimates (Mode 2) substantially above the level that could be reached with a Poisson distribution. It also increases the standard deviation above the level reached without amplification: thus, amplification does not provide a gain in angular resolution, at least for gain values below $g = 1000$: the Heisenberg uncertainty relation remains valid on the ensemble of incoming astronomical photons. For arbitrarily large gain values, we expect the standard deviation to tend towards, but never to fall below, the value reached without amplification. 

\begin{figure}
\centering
\includegraphics[width=.45\textwidth]{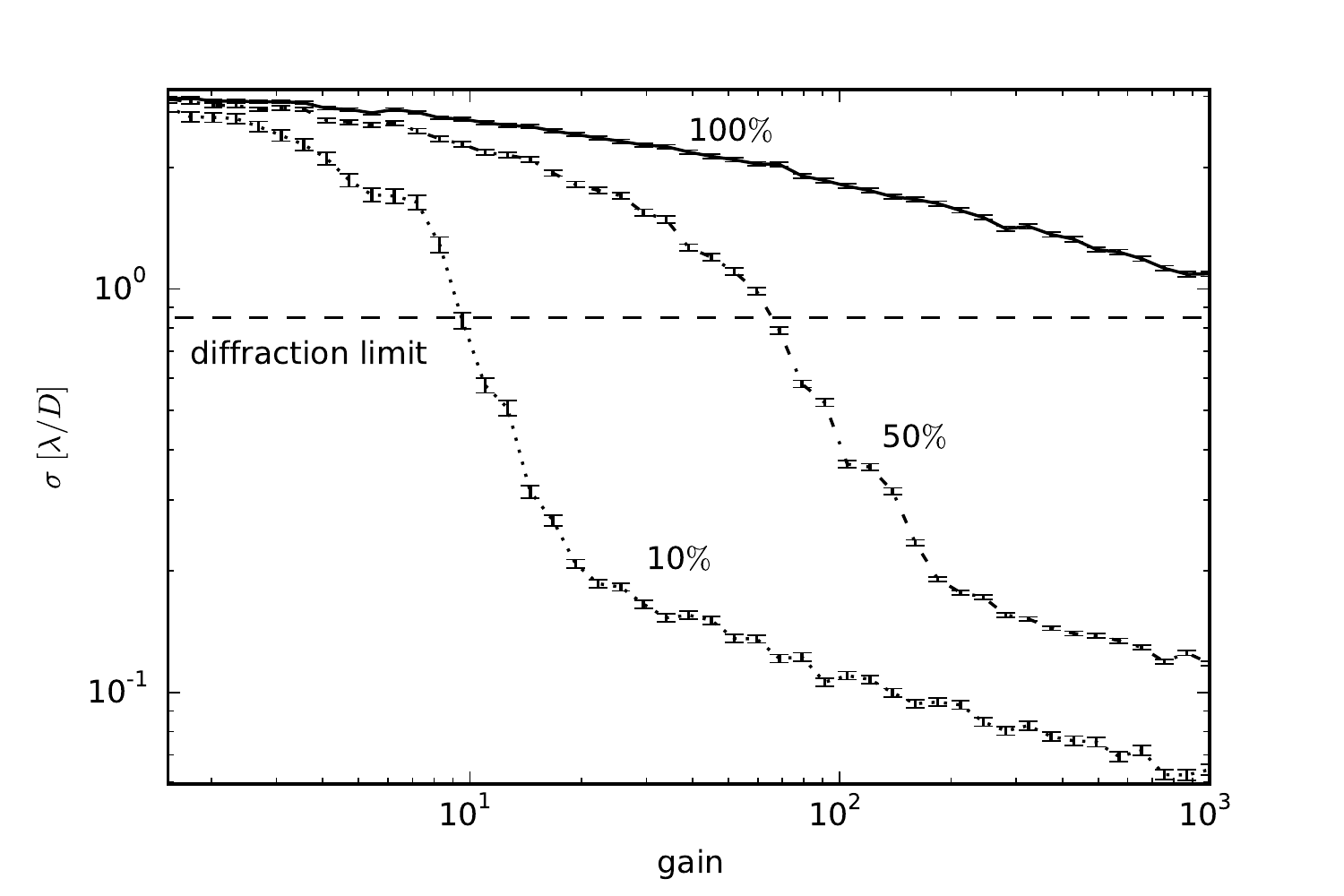}
\caption{ The rms-deviation of the position estimate as a function of amplifier gain.  The analysis is restricted to a percentage of sets with highest photon density (Modes 2, 3 and 4 in the text). }
\label{fig:selection}
\end{figure}

The analysis is now restricted to large stimulated photon bursts. To estimate the burst size, we measure $\rho$, the photon density over the circle centered on the position estimate and of diameter $\theta_0/2$. The value of $\rho$ is saved after each iteration and the iterations are  sorted by order of increasing photon density, $\rho$.  
Two further modalities are then tested:
\begin{enumerate}
\setcounter{enumi}{2}
\item The same computations as in Mode 2, but the analysis is restricted to iterations with the highest 50\% density values.
\item The same computations as in Mode 2, but the analysis is restricted to iterations with the highest 10\% density values.
\end{enumerate}

Assume we want an angular resolution equal to half the diffraction limit: Fig.\,\ref{fig:selection} shows that 90\% of events need to be discarded if the amplification gain equals 10, while 50\% of events can be kept for a gain $g=100$.  
\begin{figure}
\centering
\includegraphics[width=.45\textwidth]{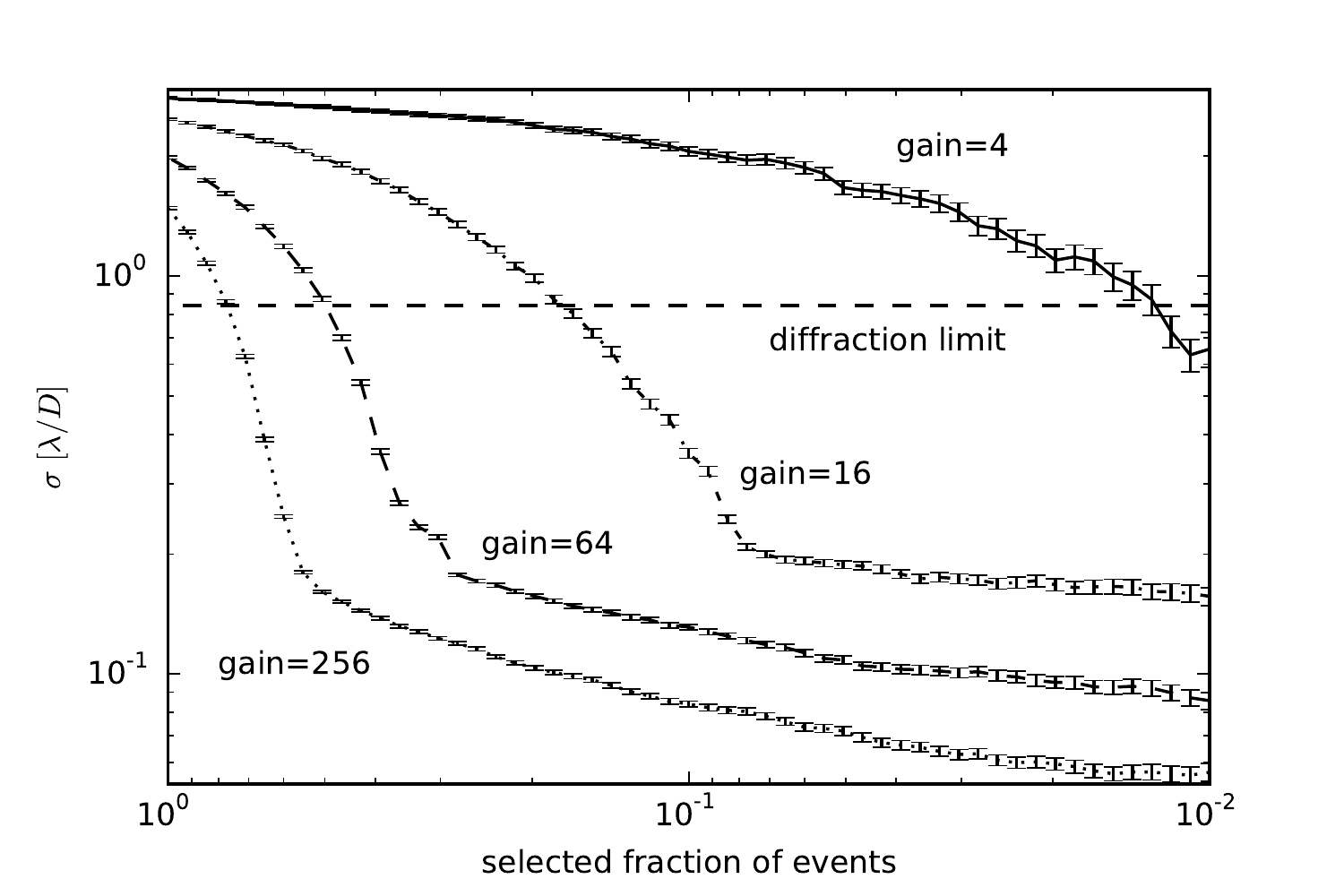}
\caption{ The rms-deviation of the position estimate as a function of the fraction of events kept for the analysis, and for different amplifier gains.  }
\label{fig:fraction}
\end{figure}

Finally, Fig.\,\ref{fig:fraction} shows the improvement in the rms-dispersion of the position estimate as a function of the fraction of events kept for the analysis. For each gain value we perform $10^4$ iterations, so that the rms-dispersion is still estimated on 100 iterations when 1\% of events are kept.  As the selection criterion becomes more restrictive, the number of stimulated photons per burst is higher and the rms-dispersion decreases:  with an amplifier gain $g = 64$ the resolution decreases to about a tenth of the diffraction limit, if only the 1\% best events are kept.

The error bars on Figs.\,\ref{fig:distr}, \ref{fig:selection} and \ref{fig:fraction} correspond to the standard error of the rms-deviation estimates. These account only for the uncertainty introduced by the finite number of iterations in our simulations. The precision of the centroiding algorithm could be improved by use e.g. of a matched filter. The rms-deviations of the position estimates can thus probably be reduced. The error bars do not account for this.  The aim of these initial simulations is to prove that it is possible to overcome the diffraction limit via optical amplification.   More refined simulations will need to explore the maximum possible gains in angular resolution. 

In conclusion, we have discussed the possibility to improve the angular resolution of a telescope through optical amplification. We extend the results recently presented by Kurek et al.\,\cite{Kurek}. With the actual distribution of stimulated photon numbers, optical amplification improves the angular resolution only if the analysis is restricted to a fraction of successful events -- that is, to a sub-set of events for which the number of stimulated photons is high. 

We suggest to estimate the number of stimulated photons from the density of photons around the final position estimate. The analysis can then be conveniently restricted to the events with high ratio of stimulated to spontaneous photons. A fraction of events is discarded, i.e. the sensitivity is reduced, but the angular resolution is improved on the remaining events with large numbers of stimulated photons. 

This method is not inherently restricted to narrow spectral bands. Photons can be amplified within broad spectral bands and QND measurements are currently\,\cite{Reiserer}, but not fundamentally restricted to narrow spectral bands. Obviously the proposed method will be especially useful for astronomical observations, if it is not restricted to narrow spectral band-widths. In microscopy the band pass limitation is not as severe.

Finally, the photon statistics behind an amplifier suggest that the quantum non-demolition (QND) stage may not be a necessity. Indeed, spontaneous photon numbers are Poisson distributed and their variance is thus fairly small. An astronomical photon that has generated a sufficiently large burst is recognized by a large photon density, several $\sigma_{\rm sp\/}$ above the average signal from spontaneous photons. Astronomical photons that generate small bursts are discarded, but the angular resolution is improved on the remaining, successfully amplified photons.  The set-up will be greatly simplified without a QND stage. We plan to investigate this possibility further.

{\bf Acknowledgments:\/} The authors gratefully acknowledge the American Astronomical Society for supporting this research with a Chr\'etien International Research Grant. 

\bibliographystyle{ieeetr}
\bibliography{AK-ER-Ref}

\end{document}